# Dietary Restriction of Amino Acids for Cancer Therapy

Jian-Sheng Kang[*]

Clinical Systems Biology Laboratories, The First Affiliated Hospital of Zhengzhou University, Zhengzhou 450052, China

[*] Correspondence should be addressed to J.-S. Kang, Clinical Systems Biology Laboratories, The First Affiliated Hospital of Zhengzhou University, Zhengzhou, 450052, China. E-mail: kjs@zzu.edu.cn ORCID ID: 0000-0002-2603-9718.

## ABSTRACT

Biosyntheses of proteins, nucleotides and fatty acids, are essential for the malignant proliferation and survival of cancer cells. Cumulating research findings show that amino acid restrictions are potential strategies for cancer interventions. Meanwhile, dietary strategies are popular among cancer patients. However, there is still lacking solid rationale to clarify what is the best strategy, why and how it is. Here, integrated analyses and comprehensive summaries for the abundances, signalling and functions of amino acids in proteomes, metabolism, immunity and food compositions, suggest that, intermittent fasting or intermittent dietary lysine restriction with normal maize as an intermittent staple food for days or weeks, might have the value and potential for cancer prevention or therapy. Moreover, dietary supplements were also discussed for cancer cachexia including dietary immunomodulatory.

**INTRODUCTION**

Cancer is a complex disease. There are more than 100 distinct types of cancer but sharing common hallmarks, including sustaining proliferative signaling and evading growth suppressors [1,2]. The anabolic and catabolic metabolisms of cancer cells must be reprogrammed to maintain their proliferation and survival, and may even hijack normal cells to create tumor microenvironment (TME) for tumorigenesis and avoiding immune destruction [2]. Due to the demands of cell growth and the needs of newly synthesized proteins as direct effectors of cellular activities, protein biosynthesis is the most energy-demanding process that accounts for ~33% of total ATP consumption [3,4,5]. Interestingly, cumulating research findings have demonstrated that amino acid (AA) restrictions play roles in cancer interventions, including glycine restriction [6], serine starvation [7,8,9], leucine deprivation [10], glutamine blockade [11,12], asparagine [13] and methionine [14]. These findings inspire and motivate a number of questions. Is there a common and effective metabolic intervention for cancer? For amino acids (AAs), which is the most heavily used AA *in vivo*? Which AA restriction is cell proliferation the most sensitive to? What kind of dietary strategies are practically available for cancer control?

**AA metabolism is the leading energy-consuming process**

The consumption and release profiles of 219 metabolites gave us glimpses of the anabolic and catabolic features of the NCI-60 cancer cell lines [6]. The consumption profiles of cancer cells represented the homogeneous demands for energy metabolism and protein synthesis, which are vital biological processes for the malignant proliferation of cancer cells. The leading substrates consumed in cancer cells included glucose and AAs, such as tryptophan, tyrosine, phenylalanine, lysine, valine, methionine, serine, threonine, isoleucine, leucine and glutamine [6]. Meanwhile, the releases features of the NCI-60 cancer cell lines showed the nonhomogeneous catabolism of glycolysis and tricarboxylic acid (TCA) cycle [6].

Another common feature of the NCI-60 cancer cell lines was the releases of nucleotides and nucleobases [6], which demonstrated that cancer cells did not directly consume nucleobases and nucleotides for their anabolism. RNA/DNA synthesis is the second energy-consuming process that contributes approximately 25% of total ATP consumption [3,4]. It is well known that cells can use AAs (including asparate, glutamine, serine and glycine, Figure 1A) as carbon and nitrogen resources for the syntheses of nucleobases [15,16,17]. Consequently, the AA metabolisms of cancer cells could use ~33-58% of the total energy expenditure (ATP) for protein synthesis and RNA/DNA synthesis (Figure 1B). From the system point of view, the energy expenditure of AA metabolisms might sufficiently underline the important and potential roles of AA restrictions in cancer interventions.

**Leucine is the most heavily used AA in human proteome**

Since the consumptions and metabolisms of AAs are the most demanding biological processes for cancer cell growth, the most heavily used AA for protein synthesis might be a potential candidate for dietary restriction in cancer therapy. Thus, the percentages of twenty AAs for all proteins in human proteome (Uniport: UP000005640) were counted, sorted and plotted from the largest to the smallest percentages for comparison (Figure 2A). As demonstrated, leucine is the most heavily used AA, serine is the second ranked, and tryptophan is the least used AA in the human proteome (Figure 2).

Leucine is an essential AA (EAA), so that it is somewhat out of expectation that leucine ranks first. Leucine is not only necessary for protein synthesis, and also acts as a signaling molecule activating mechanistic target of rapamycin (mTOR) signaling through sestrin-2 that is a cytosolic leucine sensor [18,19,20,21]. Interestingly, EAAs may ideally act as a key signal for amino acid availability. Specifically, leucine is one of branched chain AAs (BCAAs), which are not first catabolized in the liver due to the low activity of BCAA aminotransferase [22]. Consequently, leucine increases rapidly in circulation after meal [22], and are readily available as an essential nutritional signal to reduce food intake via mTOR-dependent inhibition of hypothalamic Agouti-related protein (*Agrp*) gene expression [23,24].

Although leucine is the highest enrichment AA in proteins, leucine deprivation showed modest effects on human breast cancer cells [10]. The exceptional abundance of leucine in human proteome might partially explain its limited effects on cell proliferation since leucine in degraded proteins may efflux from lysosome to meet the demand of growth [25].

### Serine is the second frequently used AA

Serine is a non-essential AA (NEAA) and ranks second. Serine metabolism is altered and enhanced in cancer cells [26,27,28]. The enhanced serine synthesis pathway could make significant contribution (~50%) to the anaplerosis of glutamine to α-ketoglutarate for mitochondrial TCA cycle [26]. Serine starvation induced stress and promoted p53-independent and p53-dependent metabolic remodelling in cancer cells [7]. Under the starvation of serine, the upregulation or enhancement of *de novo* serine synthesis pathway and oxidative phosphorylation were independent of p53, while the inhibition of nucleotide synthesis was dependent of p53-p21 activation so that the limited amount of *de novo* serine was shunted to glutathione production for the survival of cancer cells [7]. Therefore, serine starvation might have a potential role in the treatment of p53-deficient tumors.

Although glycine might be important for rapid cancer cell proliferation by supporting *de novo* purine nucleotide biosynthesis [6], glycine restriction alone didn't have the same detrimental effect on cancer cells as serine starvation, which might be explained by the inter-conversion between serine and glycine in one-carbon metabolism by serine hydroxymethyl transferase (SHMT) [7,15,16], especially the mitochondrial glycine synthesis enzyme SHMT2 [6]. Interestingly, the consumption and release profiles of the NCI-60 cancer cell lines demonstrated that glycine among AAs had the most heterogeneous pattern either consumed or released, whereas serine showed relatively homogenous consumption [6]. Beyond the potential role of glycine restriction in blocking the rapid growth of certain cancer cells, the dietary supplement of glycine was also reported to inhibit the growth of certain types of tumors, such as liver tumors [29] and melanoma tumors [30]. Therefore, the heterogeneous metabolism of glycine in cancer cells might account for its paradoxical effects.

**Lysine is a particularly important EAA**

NEAA restrictions play limited roles in cancer therapy since there are *de novo* synthesis pathways, such as serine [7]. Therefore, EAAs were focused and plotted with a log scale so that we could have a close view of EAA enrichment profiles (Figure 2B). Lysine is particularly noteworthy (Figure 2B). The AA compositions (represented in percentages) of fourteen plants and nine animals were analyzed, and the AA medians and standard deviations (SD) of these proteomes were plotted and compared with human AA abundances (represented as the width of bubbles in Figure 2C). The relative abundances of AAs in plants, animals or human were almost identical with subtle differences, and lysine ranked in the middle upper level of AAs (Figure 2C).

Those EAAs rich proteins might be particularly vulnerable to EAA restrictions, so that the 3-sigma upper limit (median + 3x SD) was applied to sort EAA exceptional rich proteins (ERPs). The numbers of ERPs rang from 604 to 1298 for individual EAA. The averaged percentages (in median ± SD) of lysine, valine and leucine in human proteome were 5.26% ±

3.27%, 5.88% ± 2.50% and 9.9% ± 3.71% respectively (Figure 2A). In comparison, the averaged percentages (in median ± SD) of lysine ERPs (n = 918), valine (n = 595) and leucine ERPs (n = 604) were 16.99% ± 3.65%, 14.89% ± 2.62% and 23.28% ± 4.5% respectively (Figure 2B). Interestingly, although the abundance of lysine in human proteome ranked third after leucine and valine (Figure 2A and 2C), the abundance of lysine ERPs ranked second only after leucine (Figure 2B). The lowest averaged percentage of ERPs was tryptophan ERPs (6.05% ± 1.86%, n = 1298) (Figure 2B).

ERPs were further chosen for functional enrichment analysis with the web server of g:Profiler [31]. To our surprise, none term of gene ontology – molecular function (GO-MF) was met the significant threshold ($p < 0.001$) for leucine, whereas lysine ERPs showed a number of enriched GO-MF terms including DNA/RNA/chromatin binding, structural constituent of ribosome, nucleosome binding and so on (Figure 2D). Together, these suggested that lysine and its ERPs were very important for cell functions, and that cell growth might be particularly vulnerable for lysine restriction. Indeed, lysine deprivation could completely block the proliferation of either p53-competent or p53-deficient cancer cells [7].

**Lysine deficiency related disease: kwashiorkor**

Kwashiorkor is a severe protein malnutrition disease of childhood associated with lysine deficiency in normal maize diet [32]. Normal maize has more protein than rice but containing low levels of two EAAs - lysine and tryptophan, which lead to the imbalance of amino acids and malnutrition [32,33]. Nowadays, maize as one of daily staples is biofortified and named as quality protein maize (QPM). QPM contains an *opaque-2* gene. The *opaque-2* gene codes a transcriptional activator so that QPM expresses more lysine and tryptophan rich proteins [33]. In the Williams' report about kwashiorkor first published in 1933 and republished in 1983, five cases were described in detail; all cases had a history of lacking breast-feeding, and were only fed with the food prepared from normal maize (cassava also used in case 5); it took four to twelve months for the development of the kwashiorkor disease in those children [32]. According to the abundance of lysine and tryptophan as shown in Figure 2, lysine deficiency might be the leading cause for kwashiorkor since the averaged percentage of tryptophan in proteins was 1.17% ± 1.3% (median ± SD) and significantly less than the abundances of other AAs (Figure 2).

Besides the important functions of lysine ERPs and the growth-halting effects of lysine deprivation discussed above, lysine is also a versatile AA modified by various modifications including methylation, acetylation, phosphorylation, malonylation, *O*-GlcNAcylation, SUMOylation, ubiquitination and lactoylation, especially those lysine residues in histones [34,35,36]. These post-translational modifications of lysine and its ERPs regulate the structures and functions of enzymes to expand the functional proteome [36], represent the crosstalk between metabolism and epignome [37], and also link cell signaling and metabolic reconfiguration to cell proliferation and differentiation [38]. Thus, all these support that lysine is an important EAA, its ERPs and their modifications play indispensable roles in homeostasis, proliferation, differentiation and diseases including malnutrition and cancer.

**Tryptophan is the least used and available EAA**

Tryptophan is an interesting and unique AA, which is the least used (Figure 2A) and also the least available AA from animal or plant foods (Figure 2C). As an EAA, tryptophan cannot be synthesized *in vivo*, and must be acquired from foods. This particular characteristic of tryptophan gave its additional roles beyond as a necessity in protein synthesis. For instance, immune system could induce tryptophan degradation to inhibit the growth of certain cancer cells [39] via interferon gamma (IFN-γ) upregulating the tryptophan-catabolizing activity of indoleamine 2,3-dioxygenase (IDO) [40]. Meanwhile, cancer cells could use the same mechanism to impede and escape immune response in TME [41,42]. Consequently, the clinical trials of IDO inhibitors showed limitations and off-target effects due to the multifaceted tryptophan metabolism [43].

The metabolism of tryptophan might be the most complicated one among AAs, and was involved in the regulation of immunity, neuronal function and intestinal homeostasis [44]. Majority (~95%) of absorbed tryptophan degraded via kynurenine pathway, in which IDOs and tryptophan-2,3-dioxygenase (TDO) were the main rate-limiting enzymes. Besides the usage for protein synthesis, a small fraction of tryptophan was catabolized by tryptophan hydroxylase (TPH) for the production of serotonin (5-hydroxytryptophan, 5-HT) and melatonin. Tryptophan and its metabolites were used and catabolized by various organs and cells to further generate bioactive metabolites, including neuroprotective kynurenic acid by astrocytes, neurotoxic quinolinic acid by microglia, neuromodulator tryptamine, immune suppressive metabolites (such as 3-hydroxykynurenine, 3-hydroxyanthranilic acid and xanthurenic acid) [44,42]. The catabolism of tryptophan induced by IFN-$\gamma$ in cancer cells and macrophages showed that the catabolites of tryptophan differed in kynurenine, anthranilic acid and 3-hydroxyanthranilic acid [40].

Tryptophan was mainly catabolized by TDO in liver, then oxidized to acetoacetyl-CoA and used for the synthesis of nicotinamide adenine dinucleotide ($NAD^+$) [44]. Thus, TDO knockout mice showed increased levels of tryptophan in plasma and 5-HT in the hippocampus and midbrain, and hence demonstrated anxiolytic modulation and adult neurogenesis [45], which was consistent to the cumulated evidence about the role of 5-HT in neurogenesis and anti-depression recently intensively reviewed by us [46,47]. It was also well known decades ago that 5-HT was involved in food intake and mood [48], and recent research focused on the functional modulation of $5-HT_6$ receptor including its agonists and antagonists [49,50]. Both the agonists and antagonists of $5-HT_6$ receptor could reduce food intake [49,50], which suggested that the activation curve of $5-HT_6$ receptor was likely a bell shape and that a proper concentration of 5-HT might increase food intake through $5-HT_6$ receptor. In short, tryptophan metabolism and its catabolites played active roles in proliferation, immunity, neurogenesis, anxiety, depression and food intake under physiological conditions.

### AA signalling in metabolism and beyond

There are two amino acid-sensing kinases: mTOR complex 1 (mTORC1) and general control nonderepressible 2 (GCN2). The landmark findings of AA sensors and signalling were emerging in recent five years (Figure 3). Besides that sestrin-2 was identified as the leucine sensor mentioned above [18,19], progress about AA sensors was recently reviewed as a part of mTOR-mediated nutrient and growth signalling [51,20,21]. Briefly, arginine has two sensors: a cytosol sensor is cellular arginine sensor for mTORC1 (CASTOR1) [52,53], and SLC38A9 is a lysosomal arginine sensor [54,55]. methionine is indirectly sensed as S-adenosylmethionine (SAM), which is sensed by SAM sensor upstream of mTORC1 (SAMTOR) [56]. ADP ribosylation factor 1 (Arf-1) relays the glutamine signals to mTORC1 activation, which is independent of the Rag GTPases [57].

Considering the importance of lysine and its ERPs, lysine sensor might be a remaining missing piece of the jigsaw (Figure 3). Lysyl-transfer RNA (lysyl-tRNA) synthetase (KRS) might act as one of potential lysine sensors, since that KRS has noncanonical roles as a secreting or nucleus-translatable signaling molecule in immune response [58,59]. Interestingly, leucyl-tRNA synthetase (LRS) was demonstrated as an intracellular leucine sensor [60,61]. Moreover, the noncanonical activities of aminoacyl-tRNA synthetases were reported for methionyl-tRNA synthetase stimulating rRNA biogenesis [62], glutaminyl-tRNA synthetase blocking apoptosis [63], tyrosyl-tRNA synthetase [64] and tryptophanyl-tRNA synthetase [65,66] acting as cytokines.

Methionine is the penultimate EAA (Figure 2B,C). Methionine is essential for the initiation of protein synthesis, while N-formylmethionine-tRNA generated in mitochondrial folate cycle is the initiator of mitochondrial protein synthesis [67] (Figure 3). SAM provides the methyl group for epigenetic modification. The methionine cycle coupling with mitochondrial energy metabolism produces SAM, so that SAMTOR might sense not only methionine, also one-carbon resource and energy levels. The methionine cycle and folate cycle are two functional modules connected and involved in one-carbon metabolism [15,16,68]. Serine acts as one-carbon donor and tetrahydrofolate (THF) serve as one-carbon receptor in the folate cycle [16]. In the methionine synthase reaction (vitamin $B_{12}$ as a essential cofactor), methyl-THF donates its methyl group via vitamin $B_{12}$ to homocysteine to produce methionine and THF [69]. The one-carbon metabolism of cancer cell may mobilize multiple carbon sources including glucose, serine, threonine, glycine, formate, histidine and choline [68]. The syntheses of pyrimidine and purine nucleotides require carbon and nitrogen sources, which rely on AA metabolism and the folate cycle in accordance with the energy metabolism and activities of mitochondria [17]. In the folate cycle, THF is an essential factor for nucleotide synthesis and the survival of cancer cells, and formyl-THF serves as one-carbon reserve [68]. Clearly, one-carbon metabolism is essential for multiple physiological processes (Figure 3) including nucleotide metabolism (especially purine synthesis), glutathione (synthesized from glutamate, cysteine and glycine) and NAPDH synthesis for antioxidant defense [17,68]. Consequently, one-carbon metabolism and nucleotide metabolism were altered and involved in the effects of the restrictions of glycine, serine and methionine [6,7,14]. Interestingly, the dietary supplementation of histidine upregulated the histidine degradation pathway to deplete THF and then enhanced the sensitivity of cancer cells to methotrexate, an inhibitor of dihydrofolate reductase for THF synthesis [70].

As mentioned, there are two amino acid-sensing kinases: mTORC1 and GCN2. The availabilities of AAs are perceived by their sensors for the activation of mTORC1 signalling. Whereas, AA deficiency is sensed by GCN2 through binding to uncharged tRNA (Figure 3). GCN2 remains phosphorylated under normal condition and dephosphorylated under AA starvation with a higher affinity to tRNA [71]. Consequently, GCN2 activation by uncharged tRNA represses the translation of most mRNAs via the phosphorylation of eukaryotic initiation factor 2 alpha (eIF2$\alpha$) but selectively increases the translation of activating transcription factor 4 (ATF4, its yeast ortholog GCN4) [72,73]. ATF4, a transcriptional master regulator of AA metabolism and stress responses (Figure 3), can upregulate AA metabolism including AA synthetases (including asparagine and serine), transporters (such as glutamine) and sensors for cancer cell proliferation via the cooperation of KDM4C-mediated H3K9 demethylation and ATF4-mediated transactivation [74,75]. Under AA deprivation, GCN2 induces the expression of sestrin-2 via ATF4 for sustaining the inhibition of mTOR signalling[74].

### AA metabolism in immune evasion and response

As mentioned in the tryptophan section, the deprivation of the lowest available EAA - tryptophan (Figure 2) was utilized either by immune system for defense or by cancer cells for immune escape in TME. In addition to IFN-γ upregulating the tryptophan-catabolizing activity of IDO [40], transforming growth factor beta (TGF-β) reduced nitric oxide synthase (NOS) and stimulated the activity of arginase 1 (ARG1) in macrophages [76]. The abnormal catabolism of tryptophan and arginine is a common TME hallmark [42]. The depletion of tryptophan[77] or arginine[78,79] led to the proliferative arrest, anergy induction or apoptosis of T cells via GCN2 activation [78,80]. Arginine and tryptophan catabolism were functionally related in IFN-γ-primed macrophagocytes. Arginine catabolism decreased the consumption of tryptophan through that nitric oxide generated by NOS inhibits the activity of IDO in mononuclear cells or monocyte-derived macrophages [81]. Cancer cells with the expression of IDO recruited and activated myeloid-derived suppressor cells (MDSCs) in TME for immune suppression [82]. MDSCs with high ARG1 activity depleted arginine and impaired T cell proliferation and cytokine production accounting for tumor evasion [79].

Interestingly, increasing the extracellular tryptophan concentration resulted in a linear induction of tryptophan catabolism by macrophages, but significantly decreased the tryptophan catabolism in most of cancer cells [40]. Whereas, extracellular tryptophan didn't affect the metabolism of dermal fibroblasts, which was consistent to the level of tryptophan as the least used AA (Figure 2). Among those immune suppressive metabolites of tryptophan mentioned in tryptophan section, xanthurenic acid can inhibit sepiapterin reductase to block the synthesis of tetrahydrobiopterin (BH4), which is an essential cofactor for aromatic amino acid hydroxylases and NOS, and also required for T cell proliferation [83]. In conclusion, these findings together might suggest that dietary supplements of tryptophan and arginine with BH4 have potential to enhance anticancer immunity.

Particularly, besides the induction of tryptophan degradation for cell growth inhibition, IFN-γ inhibited tumor cell growth more extensively by the depletion of $NAD^+$ via activating poly (ADP-ribose) polymerase family member 1 (PARP1), which was also known as adenosine diphosphate ribosyl transferase (ADPRT) [39]. Both inhibitory mechanisms induced by IFN-γ appeared to be sensitive to reactive oxygen species (ROS) and inverse proportional to glutathione concentrations [39]. These might suggest that dietary restrictions of serine, glycine, cysteine or glutamate might enhance the inhibitory effect of IFN-γ via decreasing intracellular glutathione, which was matching that manipulating ROS could modulate the responses of tumors to ROS and the survival of cancer cells under serine and/or glycine starvations [7,9]. Moreover, the dietary supplement of tryptophan might enhance the proliferation of immune cells but not much affect the inhibitory effect of IFN-γ especially via $NAD^+$ depletion in cancer cells.

**Dietary strategies for cancer therapy**

From the standpoint of energy expenditure and anabolism for cell growth, AA restriction might be an effective metabolic intervention for cancer since the energy expenditure (ATP) of AA metabolism is up to ~58% for protein and nucleotide syntheses (Figure 1). Although leucine is the most heavily used AA in human, plants and animals (Figure 2), lysine rather than leucine is likely the AA which deprivation cell proliferation is the most sensitive to. At cell and disease levels, the magnitude of lysine value was demonstrated by that lysine restriction completely blocked the proliferation of cancer cells [7], and that lysine deficiency caused human childhood malnutrition disease - kwashiorkor [32]. Tumor progression represents the survival and outgrowth of cancer cells from the battle with normal cells and immune system in TME. It is possible to gain time for tipping the scale of the battle in favor of anticancer immune responses by an effective metabolic intervention to inhibit the proliferation of cancer cells. For lysine restriction, it might be safe to be applied intermittently for days or weeks, even one or two months since it took 4-12 months for the disease development of kwashiorkor. Consequently, intermittent dietary lysine restriction might have the value and potential as a practically available dietary strategy for cancer therapy.

 **AA compositions in foods**

From a practical perspective, valuable dietary intervention using daily foods is great convenient for cancer prevention or therapy. Therefore, AA compositions of foods were collected from the website of the Food and Agriculture Organization (FAO) of the United Nations, including "Amino-acid content of foods and biological data on proteins" [84] and

"amino acid composition of feedstuffs available in the Philippines" of "the table 46 in Aquaculture feed and fertilizer resource atlas of the Philippines" [85]. The FAO information including ~300 foods was integrated, summarized and represented as a heatmap for the visualization and comparison of AA abundances in 11 FAO food terms (Figure 4A), where the green represented low abundances of AAs while the red represented the enrichment of AAs. The AA heatmap has unambiguously demonstrated that some FAO terms almost all contain low level of protein, including starch roots and tubers, vegetables or fruit (Figure 4A and 4B).

**Practical dietary AA restriction**

Due to the deficiency of lysine and tryptophan, nowadays normal maize is mainly used as feedstuffs, such as in Philippines. In Figure 4B, QPM (maize *opaque-2*) is used as a reference for AA abundances (Figure 4B). Normal maize from Philippines, China and USA are all low in lysine which abundance is about one-sixth of the value in QPM (Figure 4B), and tryptophan abundance in normal maize is about half of the value in QPM. Considering the potential application of intermittent lysine restriction in cancer prevention or therapy, normal maize might have medical value as an intermittent staple food for days or weeks for lysine restriction meanwhile avoiding protein malnutrition. In addition, some common starchy foods, vegetables and fruit in these FAO terms are chosen and recommended in Figure 4B, which can serve as complementary foods to meet daily micronutrient needs and for a rich and varied diet.

Besides the intermittent dietary restriction of lysine with normal maize, intermittent globe AA restriction is also practically feasible and beneficial by well-documented intermittent fasting [86,87,88,89], or using those common foods in Figure 4B except cereals and grain products with great cautiousness for protein malnutrition. As summarized above, different AA restrictions have various effects on cancer cells including proliferation inhibition, metastasis suppression (such as asparagine restriction) [13], sensitivity to chemotherapeutic drugs and immune escape. On the other hand, various AA supplements also have beneficial effects on cancer care, like glycine supplemented for inhibiting certain cancer cells (including melanoma and liver tumors) and dietary supplementation of histidine for increasing sensitivity to methotrexate. Dietary supplements of tryptophan and arginine with BH4 might enhance anticancer immunity.

Cancer cachexia, a complex metabolic syndrome, was estimated to account for ~20% of the death of cancer patients especially when there was 30% weight loss [90]. It was estimated that ~50-86% of cancer patients would eventually suffer a syndrome of cachexia with anorexia [91]. The levels of AAs were low in cancer cachexia [92,93]. A standard treatment of patients with cancer cachexia remains elusive [94], and nutritional interventions lacking standard criteria showed heterogenous and limited beneficial clinical outcomes but some improvements to quality of life [95]. With a bold guess, cancer cachexia is likely a severe type of global AA restriction suffered for a long period, which might be an ultimate strategy used by the own organism of patients to fight cancer. Based on the summarized AA signalling and metabolism, dietary supplements of arginine, methionine and lysine might be particularly important and necessary to stimulate protein synthesis via activating mTOR signalling for cancer cachexia. The supplement or restriction of leucine need careful deliberation because that leucine acts as a key inhibitory signal for food intake, plays an indispensable role in activating mTOR signalling, and is the most enriched AA in degraded and recycled proteins. A proper amount of the dietary tryptophan supplement might be helpful to improving anorexia for patients with cancer cachexia since the role of tryptophan metabolites in food intake and mood.

There are at least five types of diets popular among cancer patients, including the alkaline, paleolithic, ketogenic, vegan and macrobiotic diets [96]. As Zick *et. al.* summarized, these diets lacked solid scientific rationales, or convincing evidence for beneficial effects on cancer [96]. Cancer patients pursued these diets for a long period would suffer various nutrient insufficiencies [96], such as vitamin D and vitamin $B_{12}$, necessary for calcium homeostasis and maintaining one-carbon metabolism respectively. One common merit of these diets was that all emphasized eating vegetables, which met the comprehensive dietary guildelines of World Cancer Research Fund/American Institute of Cancer Research (WCRF/AICR) [97]. In the WCRF/AICR recommendations [97], it was advised to "eat a diet rich in wholegrains, vegetables, fruit and beans", "limit consumption of red and processed meat" and so on. Overall, the dietary strategies of intermittent lysine restriction or intermittent globe AA restriction (such as intermittent fasting) are largely consistent to these rules with extra attention for protein malnutrition. However, it remains unclear so far how much benefit of vegetables in cancer control might be contributed by its low protein abundance (Figure 4B).

**Dietary immunomodulatory**

The adage "the best defense is a good offense" also applies here: boosting immune system is a good offense while halt of cancer proliferation with AA restriction is relative defensive. It will further tip the scale of the battle to beat cancer cells if the immune system can be boosted simultaneously with effective metabolic interventions.

Immunonutrition has been investigated for decades, including arginine, glutamine, cysteine, n-3 fatty acids, nucleotides vitamins and trace elements [98]. However, according to the latest systematic review and meta-analysis [99], immunonutrition alone didn't reduce all-cause mortality of cancer patients although reduces postoperative infection complications. In 1998, Vanderhoof conjectured the immunomodulatory effects of carbohydrates, which had little attention and few literature at that time [100]. Vanderhoof pointed out that "most of the non-energy-related effects of carbohydrates can be related to short-chain fatty acid (SCFA) production". Interesting, in contrast to the inhibition of glutamine-utilized metabolism in cancer cells, activated T cells could adapt to glutamine blockade through the upregulation of acetate metabolism to maintain energy (especially ATP) and NADPH homeostasis [12]. The adaptation capability of T cells suggested that SCFAs (including acetate, propionate and butyrate) could be used for maintaining the metabolism of immune cells under AA restriction, and also provided an explanation to the immunomodulatory effects of carbohydrates, especially oligosaccharides and dietary fiber [100,101]. Most grains (including wheat, rye and barley) contain insoluble fiber, while oats mainly have a soluble fiber in the form of β-glucan [100]. These soluble or insoluble fiber can yield the high production of SCFAs (acetate ~60%) in the caecum and proximal colon [101]. β-glucans, a group of polysaccharides also rich in the cell wall of bacteria and fungi such as mushrooms, have strong immunomodulatory effects in cancer control [102,103,104]. Mushrooms rich in β-glucan also can serve as a source of dietary fiber [105] and for AA restriction (Figure 4B).

Mushrooms are used in traditional Chinese medicine (TCM) for centuries. Importantly, TCM demonstrated immunomodulatory effects in the prevention and treatment of severe acute respiratory syndrome (SARS) using herbal prescriptions based on the theory (Treatise on Exogenous Febrile Diseases and Miscellaneous Diseases) of Zhongjing Zhang, the father of TCM [106,107]. The TCM prescriptions for SARS prevention might will have therapeutical effects on cancer cachexia since those validated herbal prescriptions of TCM during 2003 SARS outbreak might effectively boost the immune system with a status of natural AA restriction. If necessary, supplement tryptophan, arginine and BH4 with immunotherapies to enhance the proliferation and function of T cells.

**Acknowledgements**

The author thanks all researchers in the related fields for their works that have made this comprehensive summary possible. The author is also grateful to the many authors for their papers that were uncited due to the limited space. The author is grateful for Cizhong Jiang and Liyong Wang's discussions about enrichment analysis. The author appreciates the support of the First Affiliated Hospital of Zhengzhou University.

**Abbreviations**

5-HT: 5-hydroxytryptophan, serotonin

AA: Amino acid

AAs: Amino acids

Arf-1: ADP ribosylation factor 1

ARG1: Arginase 1

BH4: tetrahydrobiopterin

CASTOR1: Cytosol sensor is cellular arginine sensor for mTORC1

EAA: Essential amino acid

GCN2: General control nonderepressible 2

IFN-γ: Interferon gamma

KRS: Lysyl-tRNA synthetase

LRS: Leucyl-tRNA synthetase

MDSCs: Myeloid-derived suppressor cells

mTOR: Mechanistic target of rapamycin

mTORC1: mTOR complex 1

NEAA: Non-essential amino acid

SAM: S-adenosylmethionine

SAMTOR: SAM sensor upstream of mTORC1

TPH: Tryptophan hydroxylase

IDO: Indoleamine 2,3-dioxygenase

TDO: Tryptophan-2,3-dioxygenase

TGF-β: Transforming growth factor beta

tRNA: transfer RNA

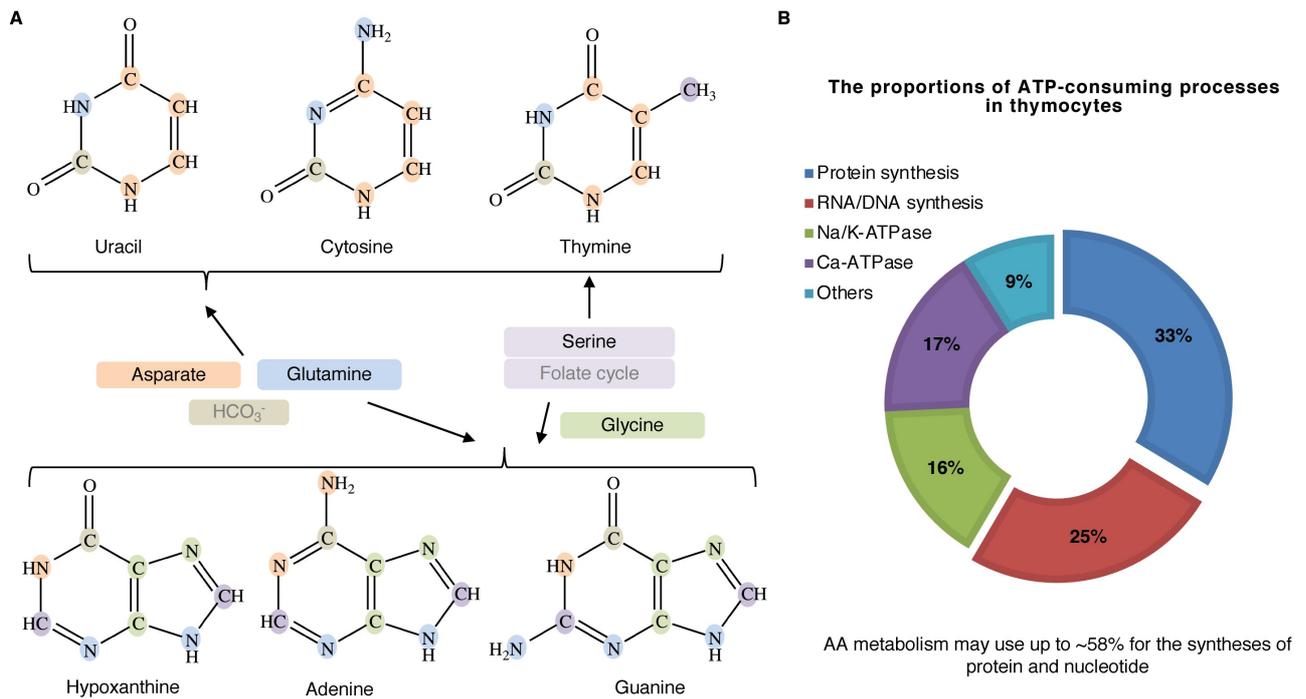

**Figure 1. Besides as constituents for protein synthesis, AAs provide carbon and nitrogen sources for the syntheses of nucleobases, and consume a large portion of ATP. A.** AAs including asparate, glutamine, serine and glycine provide the essential carbon and nitrogen sources for the syntheses of nucleobases [17]. Those carbons and nitrogens highlighted in various colors represent their original sources, such as asparate (brown), glutamine (blue), serine and folate cycle (purple), glycine (green) and bicarbonate radical (tan). **B.** The proportions of ATP-consuming processes in thymocytes [3]. AA metabolism may use up to ~58% of the total ATP consumption including protein synthesis (~33%) and nucleotide synthesis (~25%).

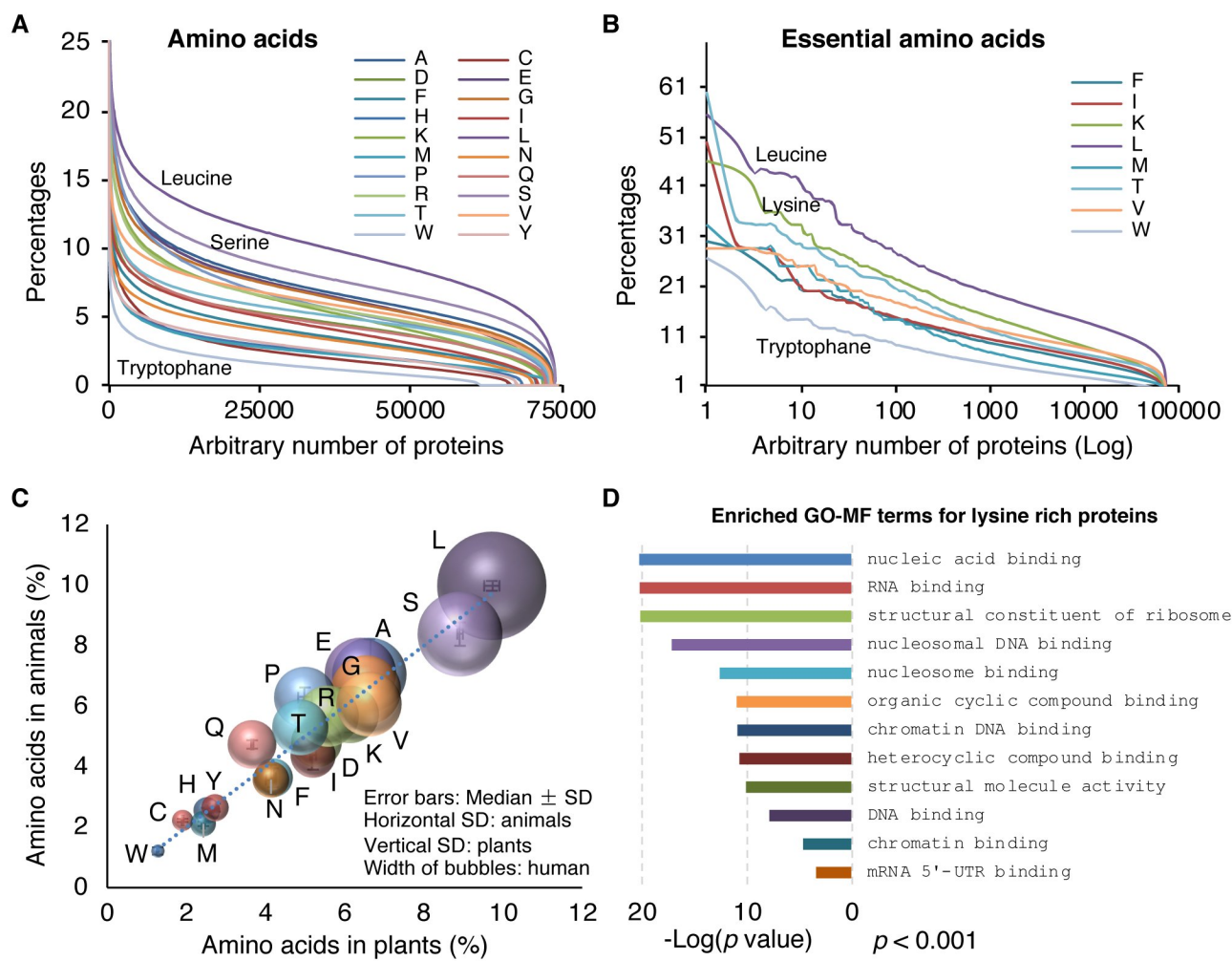

**Figure 2. A summary of AA abundances in human, plants and animals**. AAs are represented by single-letter codes: A - alanine, C - cysteine, D - aspartic Acid, E - glutamic Acid, F - phenylalanine, G - glycine, H - histidine, I - isoleucine, K - lysine, L - leucine, M - methionine, N - asparagine, P - proline, Q - glutamine, R - arginine, S - serine, T - threonine, V - valine, W - tryptophan and Y - tyrosine. **A.** Sorted percentages of AAs in human proteins. The percentages of twenty AAs for all proteins in the human proteome (Uniport: UP000005640) were sorted and plotted from the largest to the smallest percentage. **B.** Sorted percentages of EAAs in human proteins. The arbitrary number of proteins in the human proteome is plotted with a log scale for clarity, especially EAA enrichment information. **C.** A 3D-bubble plotting for AA percentages in the proteomes of human, plants and animals. human AA abundances are represented as the width of bubbles. **D.** Enriched GO-MF terms ($p < 0.001$) for lysine ERPs.

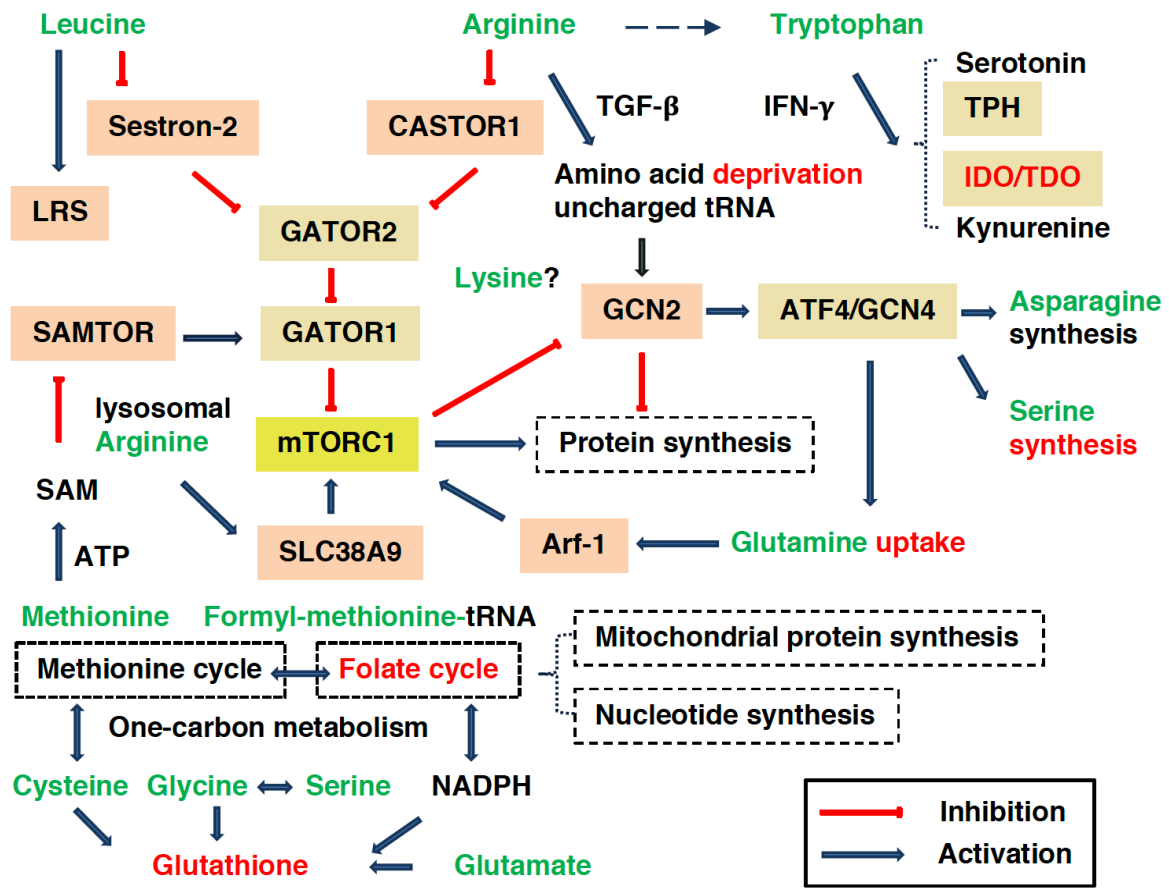

**Figure 3. A schematic of AA sensors and signalling in metabolism**. Sestrin-2 and LRS are cytosolic leucine sensors; arginine also has two sensors: a cytosol sensor - CASTOR1 and a lysosomal sensor - SLC38A9; methionine is indirectly sensed by SAM sensor – SAMTOR; Arf-1 relays the glutamine signals for mTORC1 activation. The methionine cycle and folate cycle are two functional modules connected and involved in one-carbon metabolism, which is involved in mitochondrial protein synthesis, nucleotide metabolism, glutathione and NAPDH syntheses. AA deficiency is sensed by GCN2 through binding to uncharged tRNA. GCN2 activation represses the translation of most mRNAs but selectively increases the translation of ATF4/GCN4. IFN-γ upregulates the tryptophan-catabolizing activity of IDO, while TGF-β can activate ARG1 in macrophages. The depletion of tryptophan or arginine leads to the proliferative arrest, anergy induction or apoptosis of T cells via GCN2 activation. ATF4 can upregulate AA metabolism including AA synthetases (including asparagine and serine), transporters (such as glutamine) and sensors (sestron-2) for cancer cell proliferation. Those words highlighted in red represent the upregulated biologic processes in cancer cells. LRS: leucyl-tRNA synthetase; SAM: S-adenosylmethionine; TPH: tryptophan hydroxylase; IDO: indoleamine 2,3-dioxygenase; TDO: tryptophan-2,3-dioxygenase; Arf-1: ADP ribosylation factor 1; IFN-γ: Interferon gamma; TGF-β: transforming growth factor beta.

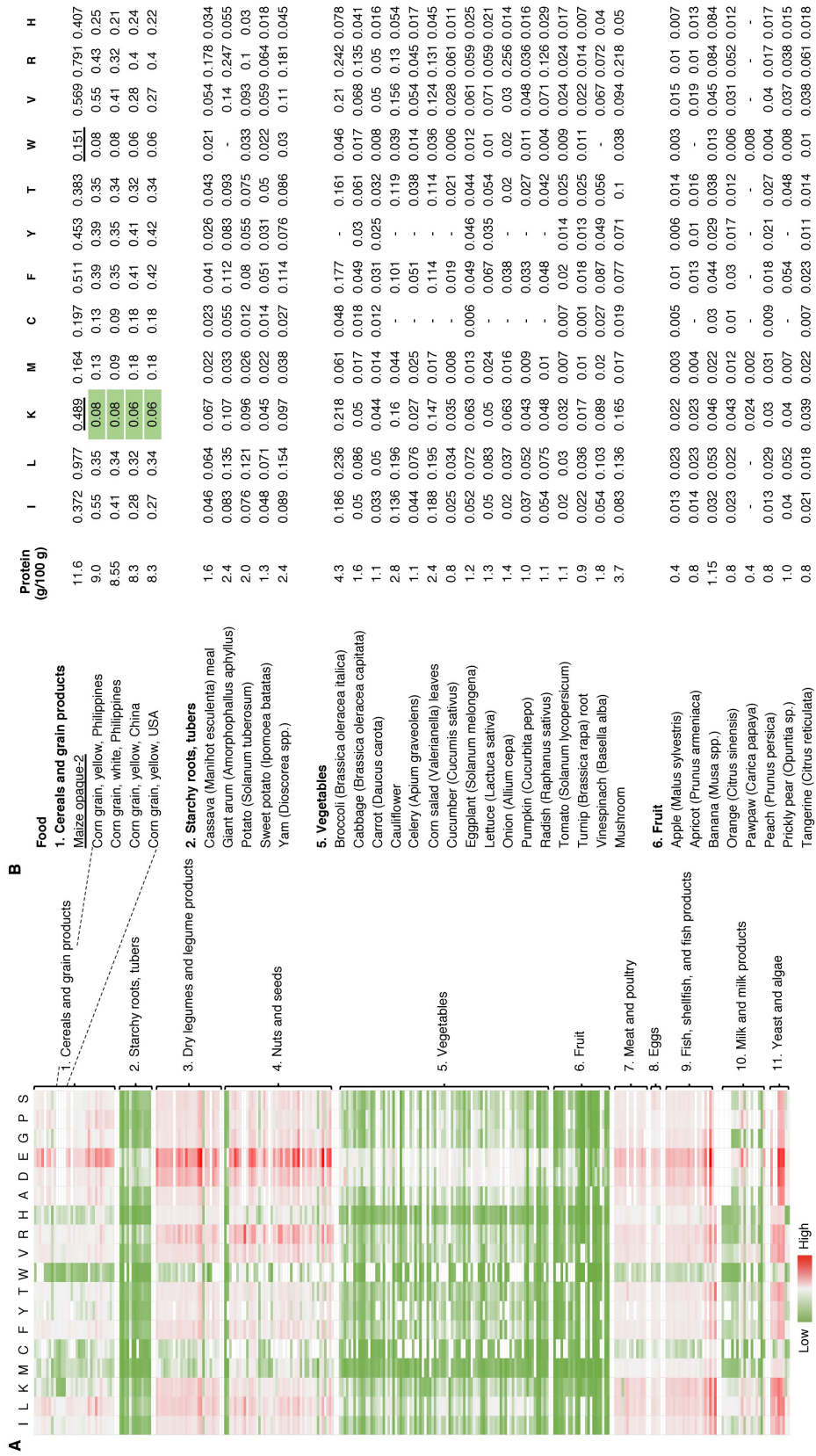

**Figure 4. AA abundances for 11 FAO terms and some common foods for AA restriction strategies**. AAs are represented by single-letter codes as Figure 2. **A.** The heatmap for ~300 foods in 11 FAO terms. The green represented low abundances of AAs while the red represented the enrichment of AAs. The AA heatmap demonstrates that some FAO terms almost all contain low level of protein, including starch roots and tubers, vegetables or fruit. **B.** Common foods chosen in those FAO terms low in protein. These foods (not a full list) low in protein are carded and recommended as samples for AA restriction strategies. QPM (maize *opaque-2*) is not a recommendation for AA restriction, but used as a reference for AA abundances. Normal maize used as feedstuffs from Philippines, China and USA are all low in lysine which abundance is about one-sixth of the value in QPM, and tryptophan abundance in normal maize is about half of the value in QPM. The listed starch roots and tubers, vegetables and fruit are all good fit for AA restriction strategies, especially mushrooms like immunonutrition.